\documentstyle[aps,twocolumn,psfig,verbatim,citesort,amstex]{revtex}
\begin{document}
\draft
\bibliographystyle{prsty}
\title{	Classical evolution of quantum elliptic states	}
\author{	Paolo Bellomo and C. R. Stroud, Jr.	}
\address{Rochester Theory Center for Optical Science and Engineering and 
The Institute of Optics, University of Rochester,
Rochester, NY 14627-0186, USA. }
\date{\today}
\maketitle
\begin{abstract}
The hydrogen atom in weak external fields is a very accurate model for 
the multiphoton excitation of ultrastable high angular momentum Rydberg 
states, a process which classical mechanics describes with astonishing 
precision.
In this paper we show that the simplest treatment of the intramanifold 
dynamics of a hydrogenic electron in external fields is based on the 
elliptic states of the hydrogen atom, i.e., the coherent states of SO(4), 
which is the dynamical symmetry group of the Kepler problem. 
Moreover, we also show that classical perturbation theory yields the 
{\it exact} evolution in time of these quantum states, and so we 
explain the surprising match between purely classical perturbative calculations
and experiments.
Finally, as a first application, we propose a fast method for the 
excitation of circular states; these are ultrastable hydrogenic eigenstates 
which have maximum total angular momentum and also maximum projection of 
the angular momentum along a fixed direction.
\end{abstract}
\pacs{32.80.Rm, 32.60.+i, 03.65.-w, 02.20.-a
-- Accepted for publication in Phys. Rev. A
}
\section{Introduction}

In the past few years innovative experimental techniques have made possible the study of 
the dynamics of ``Rydberg" electrons, that is, atomic electrons that 
are promoted to very high energy levels and that are only weakly bound to the 
atomic core {\cite{amato_96}}.
The spectrum of such electrons is well described by a Rydberg-like formula (hence 
their name), and their wave functions are well approximated by eigenfunctions of the 
hydrogen atom with very large principal quantum number 
(typically $n \gtrsim 100$) {\cite{gallagher_94}}. 
Indeed, to a very good approximation the dynamics of Rydberg electrons is hydrogenic, 
and more complex atoms are often used in experiments merely as substitutes for 
hydrogen, because it is much easier to excite their valence electron to a Rydberg 
state, and yet the far-flung Rydberg electron senses a field which does not differ 
much from a pure Coulomb field. 
The recent experimental results have led to a renewed 
theoretical interest in the hydrogen atom in external fields in the limit of large 
quantum numbers, which is an exemplar for
the study of quantum-classical correspondence in nonintegrable 
systems {\cite{gutzwiller_90}}.

Indeed, recent experiments have shown that the intramanifold 
dynamics of large-{\it n} Rydberg electrons depends strongly 
on the presence of even surprisingly weak fields.
This observation strongly suggests that it must be possible to 
manipulate accurately the quantum state of the electron 
by applying the appropriate combination of
weak, slowly varying electric and magnetic fields.
In fact, the theory of the hydrogen atom in weak fields is the basis of the treatment
of slow ion-Rydberg collisions, which are generally considered to be the mechanism
for the stabilization of the high-{\it n} states used in 
ZEKE (zero-electron-kinetic-energy) spectroscopy
{\cite{macadam_93a,myself_7,myself_8}}. 
It also constitutes the starting point for 
the study of alkali atoms in weak, circularly polarized microwave 
fields {\cite{gross_86a,gallagher_97,myself_6,myself_9,nauenberg_93a}}.

An external electric field ${\bold F}$ is ``weak" when its
magnitude is small compared to the average Coulomb field sensed by 
the electron, that is, in atomic units (which we use throughout this paper)
\begin{equation}
\label{c_1}
	F << \frac{1}{n ^{4} } 	\; .
\end{equation} 
In classical mechanics Eq. ({\ref{c_1}})
implies that the energy of the Rydberg electron does not change significantly
over a Kepler period, and classical perturbation theory applies.
The condition on a magnetic field ${\bold B}$ is that the magnitude of the 
field must be much smaller than 
the Kepler frequency $ \omega_{K} = 1 / n^{3} $ of the electron.

On the other hand, the quantum constraint on electric fields
for negligible intermanifold mixing is the 
Inglis-Teller limit {\cite{gallagher_94}}
\begin{equation}
\label{c_2}
	F < \frac{1}{ 3 n ^{5} } \; ,
\end{equation}
and for a very large {\it n} it may become a more stringent constraint 
than the classical one.
The quantum constraint on a magnetic field  $ B $ is
\begin{equation}
\label{c_2a}
	B < \frac{1}{ n^{4} } \; .
\end{equation}
However, it has been recently shown 
that the more relaxed
classical constraints hold also in quantum mechanics. 
That is, even in the presence of some intermanifold mixing the slow {\it secular} 
dynamics, due to the external fields, is essentially the same as if the Rydberg electron
were still confined within a given {\it n}-manifold, because the time-averaged
corrections due to {\it n}-mixing are negligible for large {\it n} {\cite{myself_12}}.
Therefore in this paper we consider only the intramanifold dynamics of the Rydberg 
electron, and we assume that the external fields satisfy the Inglis-Teller limit, and
also Eq. (\ref{c_2a}).

This paper is organized as follows: 
in Sec. II we discuss the evolution in time of atomic elliptic states in weak fields,
and we show that they evolve exactly like the underlying classical ellipse;
in Sec. III we propose an original approach to the production of circular Rydberg states,
which is based on the dynamics of the coherent states of $SO(4)$; finally,
in Sec. IV we draw some general conclusions.

\section{Perturbative dynamics in quantum and classical mechanics}

The Hamiltonian for a hydrogen atom in crossed electric and magnetic fields reads
\begin{equation}
\label{c_3}
	H = \frac{p^2}{2} - \frac{1}{r} + \frac{B}{2} L_{z} + Fx + \frac{ B^2 }{8}
	( x^2 + y^2 ) \; ,
\end{equation}
where the electric field is parallel to the $x$ axis and its 
strength is $F$; the magnetic field is parallel to the $z$ axis 
and its strength is $B$.

For weak fields the diamagnetic term, which is
proportional to the square of the field, can be neglected.
The simplified problem has been first solved quantum mechanically by 
Demkov {\it et al.} {\cite{demkov_70,demkov_74}}.
However, their formal solution does not provide physical 
insight in the dynamics of the angular momentum of the Rydberg electron;
it also becomes computationally intractable in the limit of large {\it n}'s.

The analysis of the intramanifold dynamics in the hydrogen atom
rests on Pauli's replacement, which is 
an operator identity between the position operator
${\bf {\hat r}}$ and the scaled Runge-Lenz vector operator ${\bf {\hat a}}$
(throughout this paper we use boldface letters for vectors, and
we indicate a quantum operator with a caret), 
and which holds only within a hydrogenic {\it n}-manifold 
{\cite{pauli_26a,englefield_72}}:
\begin{equation}
\label{c_4}
	{\bf {\hat r} } = - \frac{3}{2} n {\bf {\hat a} } \; .
\end{equation}
The scaled Runge-Lenz vector operator ${\bf {\hat a}}$ is a hermitian 
operator, which for a bound state is defined as
\begin{equation}
\label{c_5}
	{\bf {\hat a}} = \frac{1}{\sqrt{-2E}} 
        	\left\{   
                       \frac{1}{2} 
                       \left(    
                             {\bf {\hat p}} \times {\bf {\hat L}} - 
			     {\bf {\hat L}} \times {\bf {\hat p}} 
                       \right) 
                       - \frac{ {\bf {\hat r}} }{r} 
                \right\} \; ,
\end{equation}
where $ E = - 1 / 2 n^{2} $ is the Kepler energy of the electron.

The angular momentum and the Runge-Lenz vector are invariants of the 
Kepler problem, and they commute with the hydrogenic Hamiltonian. 
By neglecting the diamagnetic term and using the 
identity of Eq. ({\ref{c_4}}), in the interaction representation the 
perturbation Hamiltonian for external fields of arbitrary orientation becomes
\begin{equation}
	\label{c_6}
	{\hat H}_{1} = - {\boldsymbol \omega}_{S} \cdot {\bf {\hat a} } - 
		  {\boldsymbol \omega}_{L} \cdot {\bf {\hat L} } \; ,
\end{equation}
where $ {\boldsymbol \omega}_{S} = 3 n {\bf F } / 2 $ is the Stark frequency
of the electric field, and ${\boldsymbol \omega}_{L} = - {\bf B} / 2 $ is the Larmor
frequency of the magnetic field
(we define the vector Larmor frequency with a minus sign, so that
the dynamics is formally identical to the one of a negative charge in a 
noninertial rotating frame - see below).

The components of the angular momentum, plus those of the Runge-Lenz vector
constitute the generators of $ SO(4) $, which is the dynamical symmetry group
of the Kepler problem {\cite{englefield_72}}. 
It is convenient to decompose $ SO(4) $ in the direct product of 
two rotation groups, i.e. $ SO(4) = SO(3) \times SO(3) $, and we 
consider the following operators:
\begin{equation}
\label{c_7}
	\begin{split}
		{\bf {\hat J}}_{1} & = 
		\frac{1}{2} \left( {\bf {\hat L}} 
		+ {\bf {\hat a}} \right) \\
		{\bf {\hat J}}_{2} & = 
		\frac{1}{2} \left( {\bf {\hat L}} 
		- {\bf {\hat a}} \right)  \; .
	\end{split}
\end{equation}
It is well known that ${\bf {\hat J}}_{1}$ and ${\bf {\hat J}}_{2}$ 
commute with each other and that their components
constitute a realization of the angular momentum algebra {\cite{englefield_72}}.
The perturbation Hamiltonian can be rewritten as
\begin{equation}
\label{c_8}
	{\hat H}_{1} = - {\boldsymbol \omega}_{1} \cdot {\bf {\hat J} }_{1} - 
		  {\boldsymbol \omega}_{2} \cdot {\bf {\hat J} }_{2} \; ,
\end{equation}
where
\begin{equation}
\label{c_9}
	\begin{split}
		{\boldsymbol \omega}_{1} & = 
		{\boldsymbol \omega}_{L} +
		{\boldsymbol \omega}_{S} \\
		{\boldsymbol \omega}_{2} & = 
		{\boldsymbol \omega}_{L} -
		{\boldsymbol \omega}_{S}  \; .
	\end{split}
\end{equation}
Moreover $ {\bf {\hat L}} $ and $ {\bf {\hat a}} $ obey two constraints:
\begin{equation}
\label{c_9aa}
	\begin{split}
		& {\bf {\hat L}} \cdot {\bf {\hat a}} = 0 \\
		& {\bf {\hat L}}^{2} + {\bf {\hat a}}^{2} = n^{2} -1 \; ,
	\end{split}
\end{equation}
and so one has
\begin{equation}
\label{c_9a}
	\begin{split}
		& {\bf {\hat J}}_{1}^{2} = {\bf {\hat J}}_{2}^{2} = j ( j + 1 ) \\
		& 2 j + 1 = n \; .
	\end{split}
\end{equation}
Therefore both irreducible representations of $ SO(3) $ have the same 
dimension, which is related to the 
principal quantum number {\it n} of the hydrogenic manifold.

Equation (\ref{c_8}) reduces the problem to the dynamics of two
uncoupled spins in the external ``magnetic fields" $ {\boldsymbol \omega}_{1} $
and $ {\boldsymbol \omega}_{2} $.
The analysis is particularly simple when the two ``magnetic fields"  
$ {\boldsymbol \omega}_{1} $ and $ {\boldsymbol \omega}_{2} $
have constant orientation in space. 
However, all the considerations below
also hold in the more general situation of {\it arbitrary}
fields within the constraints of perturbation theory
{\cite{klauder_85,perelomov_86}}, and we discuss 
explicitly the greater generality
of our analysis later in this section. 
In the case of ``magnetic fields" 
with constant orientation the propagator is simply
\begin{equation}
\label{c_10}
	{\hat U} ( t'',t' ) = 
	e ^{ 
	i \int_{t'}^{t''} {\boldsymbol \omega}_{1} \cdot {\bf {\hat J}}_{1} dt 
	} 
	e ^{ 
	i \int_{t'}^{t''} {\boldsymbol \omega}_{2} \cdot {\bf {\hat J}}_{2} dt 
	} \; .
\end{equation}

The elliptic eigenstates of the hydrogen atom 
{\cite{delande_89a,delande_90a,delande_91a,nauenberg_89a,nauenberg_94a}} are
nothing other than the coherent states of $ SO(4) $ and therefore 
they can be expressed as the direct product
of two coherent states of $ SO(3) $. 
In turn, the coherent states of $ SO(3) $ can be 
constructed quite generally by applying any operator of the group 
(i.e., any rotation) to the angular momentum eigenstate
with maximum projection of the angular momentum along 
the {\it z} axis {\cite{klauder_85,perelomov_86}}:
\begin{equation}
\label{c_11}
	| j {\bf n}_{1} , j {\bf n}_{2} \rangle
	= 
	e ^{ i 
	{\boldsymbol \Omega}_{1} \cdot {\bf {\hat J}}_{1} }
	e ^{ i 
	{\boldsymbol \Omega}_{2} \cdot {\bf {\hat J}}_{2} }
	| j j_{1,z} = j \rangle \otimes
	| j j_{2,z} = j \rangle \; ,
\end{equation}
where $ {\boldsymbol \Omega}_{1} $ and $ {\boldsymbol \Omega}_{2} $
represent 3-dimensional active rotations, which respectively 
overlap the {\it z} axis with the unit vectors ${\bf n}_{1}$ and 
${\bf n}_{2}$.

Clearly, angular momentum eigenstates 
that have maximum projection along the {\it z} axis are
minimum uncertainty states for the 
angular momentum {\cite{klauder_85,perelomov_86}}, 
and the rotations of Eq. (\ref{c_11}) preserve this property.
The coherent states of $ SO(3) $ are then states of minimum uncertainty 
for the angular momentum, and in their representation on the unit sphere they are 
sharply localized along the direction of the corresponding classical 
angular momentum. 
Similarly, elliptic states are localized along the directions of both
classical ``angular momenta" $ {\bf J}_{1} $ and $ {\bf J}_{2} $, 
i.e. along the unit vectors $ {\bf n}_{1} $ and $ {\bf n}_{2} $.
It follows from Eq. (\ref{c_7}) that
they also possess well localized, quasiclassical
angular momentum {\bf L} and Runge-Lenz vector {\bf a}.

%
%
%
%
\begin{figure}
\centerline{\psfig{file=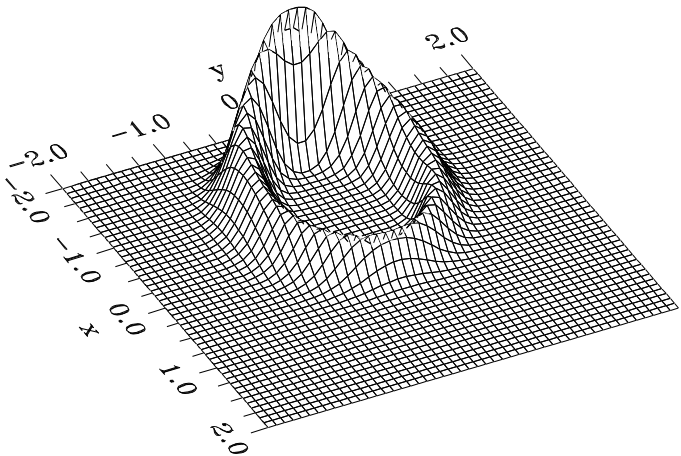,height=6.0cm,width=9.0cm,angle=0}}
\caption{
Probability density (averaged along the {\it z} axis)
of an atomic elliptic state localized on the {\it xy} plane.
The principal quantum number is $n=30$, and the eccentricity is $e_{c} = 0.6$.
Both {\it x} and {\it y} vary between $-2 n^{2}$ and $+ 2 n^{2}$ (recall that in atomic
units the Bohr radius is equal to one), and the nucleus is at the origin of the
frame of reference. 
The probability density is sharply
concentrated on the {\it xy}-plane, and we have averaged it over the {\it z} axis
to show that the electron is more likely to be far from the nucleus. 
The peak at the aphelion which reflects the larger probability of finding
the electron away from the nucleus is a purely classical effect. 
Because the classical electron is slower at the aphelion than at 
the perihelion it spends a longer time away from the nucleus than in its
proximity, and the classical probability of finding a Rydberg electron 
far from the nucleus is larger. 
The figure was produced using the 
elegant formula for the wave function 
of atomic elliptic states derived in Ref. [20].
}
\label{c_fig_1}
\end{figure}
%
%
%

The classical objects that correspond to elliptic states are 
points in the phase space of the Kepler problem.
In the more familiar configuration space these points are
the trajectories of a classical electron in a pure Coulomb 
potential, i.e., Kepler ellipses,
which are completely identified by the magnitude and direction of the 
two classical vectors $ {\bf L} $ and $ {\bf a} $ {\cite{goldstein_80}}.
Indeed, the probability density of a hydrogenic electron in an elliptic
state is peaked precisely along a 
Kepler ellipse (see Fig. \ref{c_fig_1}).

Most importantly, it is easy to see that the 
propagator of Eq. (\ref{c_10}) is also an operator of SO(4), and so
when the propagator acts on an elliptic state it 
naturally yields some other elliptic state.
More precisely, elliptic states are constructed by 
applying two rotation operators which map the 
{\it z} axis (that is, the direction of the angular momenta of the 
original states) onto some desired directions 
$ {\bf n}_{1} $ and $ {\bf n}_{2} $. 
Similarly, the propagator of Eq. (\ref{c_10}) consists of two rotations
respectively around the spatial axes given by the ``magnetic fields" 
$ {\boldsymbol \omega }_{1} $ and $ {\boldsymbol \omega}_{2} $.
The net effect of the propagator onto an elliptic state is:
\begin{equation}
\label{c_12}
	{\hat U} ( t'' , t' ) 
	| j {\bf n}_{1} , j {\bf n}_{2} \rangle =
	| j {\bf n'}_{1} , j {\bf n'}_{2} \rangle \; ,
\end{equation}
where the two final unit vectors $ {\bf n'}_{1} $, $ {\bf n'}_{2} $
can be obtained from the initial ones by a clockwise {\it classical} precession
around $ {\boldsymbol \omega }_{1} $ and $ {\boldsymbol \omega}_{2} $. 

The original idea of studying the dynamics of elliptic states in weak 
fields is due to Nauenberg {\cite{nauenberg_94a}}, 
who treated in detail the case of orthogonal,
time-dependent electric and magnetic fields. 
In his analysis the 
connection with classical mechanics emerges for a configuration 
of the fields which constitutes a realization of $ SO(3) $. 
We generalize his results to 
arbitrary fields, that is, to realizations of $ SO(4) $, the full
dynamical symmetry group of the Hydrogen atom.
Indeed, most of the interest concerning elliptic states has been focused on the 
{\em quasiclassical localization} properties of the electron in real space 
{\cite{delande_89a,delande_90a,delande_91a,nauenberg_89a,nauenberg_94a}},
and the representation of elliptic states as a direct product
of two coherent states of $ SO(3) $ 
(which yields their time dependence so naturally and so generally) 
is not particularly suited for the study of the probability density of 
Rydberg electrons in real space.

In general, the classical object which corresponds
to a coherent state is the classical phase space point which 
labels the state itself
{\cite{klauder_85,perelomov_86}}.
A point in phase space is equivalent to 
a trajectory of the electron, 
which in the case of the Coulomb potential is a Kepler ellipse.
Therefore the classical counterpart to the motion of the 
coherent states of $ SO(4) $ is the dynamics of classical ellipses in weak fields,
which is the object of study
in classical perturbation theory, where the ellipse becomes the 
dynamical object itself {\cite{born_60,percival_79,hezel_92a,myself_7}} . 

In classical perturbation theory it is assumed that the electron still moves
along an unperturbed ellipse, and the equations of motion 
describe how the elements (in the sense of celestial mechanics {\cite{szebehely_67}}) 
of the ellipse slowly vary in time. 
Clearly, an ellipse can be described by several
equivalent sets of elements, however, if one chooses the 
angular momentum ${\bf L}$ and the Runge-Lenz vector ${\bf a}$
(the magnitude of the latter being 
proportional to the eccentricity of the ellipse) the equations of motion 
turn out to be particularly simple.
More formally, in classical mechanics the angular momentum and the Runge-Lenz
vector are constants of motion for the pure Kepler problem, i.e., 
their Poisson brackets with the Hamiltonian vanish, just like the commutators
of the corresponding quantum operators. 
However, {\bf L} and {\bf a} become time-dependent
as soon as applied external fields 
break the SO(4) symmetry of the Hamiltonian.
In the case of very weak fields the effects of the perturbation take place on a 
time scale much longer than the Kepler period $ T_{K} = 2 \pi n^{3} $.
By simply averaging the equations of motion 
{\it over a Kepler period} and {\it along an unperturbed Kepler ellipse} 
one can easily derive the dynamics for the {\it time-averaged} angular momentum 
and the {\it time-averaged} Runge-Lenz vector, which for the sake of simplicity
we still indicate with $ {\bf L} $ and $ {\bf a} $, and 
one has {\cite{born_60,percival_79,hezel_92a,myself_7}}:
\begin{equation}
\label{c_13}
	\begin{split}
		\frac{ d {\bf L} }{ dt } & = 
		- {\boldsymbol \omega}_{S} \times {\bf a} 
		- {\boldsymbol \omega}_{L} \times {\bf L} \\
		\frac{ d {\bf a} }{ dt } & = 
		- {\boldsymbol \omega}_{S} \times {\bf L} 
		- {\boldsymbol \omega}_{L} \times {\bf a} \; .
	\end{split}
\end{equation}
Like in quantum mechanics, the dynamics is particularly straightforward 
when it is expressed 
in terms of the ``angular momenta" $ {\bf J}_{1} $ and $ {\bf J}_{2} $, 
which obey simple, uncoupled equations:
\begin{equation}
\label{c_14}
	\begin{split}
		\frac{ d {\bf J}_{1} }{ dt } & =  
		- {\boldsymbol \omega}_{1} \times {\bf J}_{1} \\
		\frac{ d {\bf J}_{2} }{ dt } & =  
		- {\boldsymbol \omega}_{2} \times {\bf J}_{2} \; ,
	\end{split}
\end{equation}
where the two frequencies are the same as in Eq. (\ref{c_9}).
The two classical spin vectors $ {\bf J}_{1} $ and $ {\bf J}_{2} $ simply
precess clockwise around the ``magnetic fields" $ {\boldsymbol \omega}_{1} $
and $ {\boldsymbol \omega}_{2} $, just like their quantum counterparts.

This shows that elliptic states in weak fields not only 
do evolve into elliptic states, and therefore retain their 
coherence properties and their localization
along a classical Kepler ellipse, but they also evolve {\it exactly} according to
the laws of classical mechanics (in the perturbative limit).
This result has been already observed numerically and
discussed theoretically for special
fields configurations in Refs. {\cite{carlos_97c,carlos_98a}}, 
and also in Ref. {\cite{nauenberg_94a}}, where the case
of orthogonal electric and magnetic fields is discussed.
Most importantly, the same result
has been observed also experimentally {\cite{gross_86a}}.

Indeed, we have illustrated explicitly the connection 
between quantum and classical mechanics
only for a special configuration of the fields. However,
our approach is based on the dynamics of the coherent states of $ SO(4) $
and that guarantees -see below- that our conclusions hold for {\it arbitrary} fields
(within the constraints of perturbation theory).
Therefore, our study provides an analytical explanation of the 
numerical results and it also
{\it generalizes} the previous theoretical arguments 
{\cite{carlos_97c,carlos_98a,nauenberg_94a}}.
The main conclusions of our analysis do not depend
on the particular choice of external fields; instead, they
rest on the equivalence between the 
intramanifold dynamics of a Rydberg electron in weak external fields,
and the motion of two uncoupled spins in external magnetic fields, 
and also on the properties of the coherent states of the angular momentum.
It is well known that the coherent states of
$ SO(3) $ in {\it arbitrary} magnetic fields evolve
in time exactly like the corresponding classical
spin vectors {\cite{klauder_85,perelomov_86}}, and that is the reason 
why our demonstration holds for {\it arbitrary} electric and magnetic fields.
Although the arguments for the classical evolution of quantum elliptic states
hold in general, for fields with complicated time-dependence 
the explicit form of the propagator may be difficult to 
derive analytically. However, it is easy to see that it must be given 
by some combination of rotation operators.
It is easy to see that in general the Euler angles of the 
propagator for a spin in a magnetic field
obey some complicated, nonlinear differential 
equations that must be solved numerically, when
the time dependence of the field is not trivial 
{\cite{klauder_85,perelomov_86}}.
However, when a numerical treatment is necessary, 
it is clearly much simpler to solve the classical, 
linear Eqs. (\ref{c_14}). 
In fact, the classical equations yield directly the unit vectors
${\bold n'}_{1}$ and ${\bold n'}_{2}$, which label and
determine completely the 
coherent state after it has evolved in time.

The quantum propagator of Eq. (\ref{c_10})
is just the solution for a very special configuration of the external fields, 
however it is very useful because of its {\it illustrative} character,
and of its relevance to ion-Rydberg collisions, which have been investigated
experimentally.
Moreover, for such fields the classical dynamics of the unit vectors labeling the 
elliptic state becomes amenable to an exact analytical treatment and it
yields a most intuitive understanding of
the dynamics, and we exploit this final characteristic in the next section.

A slowly rotating electric field is equivalent to 
crossed electric and magnetic fields 
in the noninertial frame rotating with the field
{\cite{goldstein_80,demkov_74,myself_7}}.
Therefore our analysis explains why 
calculations based on purely classical methods
account so well for several experimental results, ranging from 
slow ion-Rydberg collisions {\cite{macadam_93a,myself_7,myself_8}}
to the dynamics of circular states in 
circularly polarized fields {\cite{gross_86a,myself_6}}, 
and to the anomalous scaling of the autoionization 
lifetimes of alkaline-earth Rydberg atoms also in 
circularly polarized microwave fields {\cite{gallagher_97,myself_9}}.
Indeed, a classical trajectory Monte Carlo simulation
based on Eqs. (\ref{c_14}) is almost equivalent to a 
quantum treatment, in which the initial state is represented
as a superposition of coherent states of $SO(4)$.
That is, an elliptic state which is localized along a classical
ellipse follows that same ellipse during its time evolution.
Clearly, the quantum state is always somewhat diffuse, which is not true
for a classical orbit. On the other hand, the overlap between two elliptic states
with different angular momentum and Runge-Lenz vector is {\cite{klauder_85,perelomov_86}}:
\begin{multline}
\label{c_14a}
	\left| \langle j {\bf n}_{1} , j {\bf n}_{2} |
	j {\bf n'}_{1} , j {\bf n'}_{2} \rangle \right| ^{2} \\
	= \left( \frac{ 1 + {\bf n}_{1} \cdot {\bf n'}_{1} }{2} \right) ^{ n - 1 }
	\left( \frac{ 1 + {\bf n}_{2} \cdot {\bf n'}_{2} }{2} \right) ^{ n - 1 } \; ,
\end{multline}
and because ${\bf n}_{i} \cdot {\bf n'}_{i} \leq 1$
in the limit of large quantum numbers elliptic states behave
more and more like sharply localized classical ellipses.

The time evolution of the classical vectors
$ {\bf J}_{1} $ and $ {\bf J}_{2} $ which describe a Kepler ellipse 
can be expressed in terms of a
classical propagator, that is
\begin{equation}
\label{c_14b}
	{\bf J}_{i} ( t'' ) = U^{\rm cl}_{i} ( t'', t' ) {\bf J}_{i} ( t' ) 
	\;\;\;\; i = 1,2 \; ,
\end{equation}
and we conclude this section by writing explicitly the classical propagator for 
the important case when $ {\boldsymbol \omega}_{1} $ and $ {\boldsymbol \omega}_{2} $ 
have constant orientation in space, that is
\begin{equation}
\label{c_15}
	{\boldsymbol \omega}_{i} = 
	\omega_{i} (t) {\bf n}_{\omega_{i}} \; , \;
	{\bf n}_{ \omega_{i} } =
	\left( 
	{\tilde \omega}_{ix} , 
	{\tilde \omega}_{iy} ,
	{\tilde \omega}_{iz} 
	\right) \;\;\;\; i = 1,2	\; ,
\end{equation}
where ${\tilde \omega}_{ix}$, ${\tilde \omega}_{iy}$ 
and ${\tilde \omega}_{iz}$ 
are the components of the unit vector ${\bf n}_{\omega_{i}}$ 
that points along ${\boldsymbol \omega}_{i}$.
First we set
\begin{equation}
\label{c_16}
	\phi_{i} = \int_{t'}^{t''} dt \omega_{i} (t) \;\;\;\; i = 1,2 \; ,
\end{equation}
and the classical propagator is
\begin{multline}
\label{c_17}
	U^{\rm cl}_{i} ( t'' , t' ) =
	\cos \phi_{i} \cdot I - \sin \phi_{i} \cdot N_{i} \\
	+ ( 1 - \cos \phi_{i} ) P_{i}
	\;\;\;\; i = 1,2 \; ,
\end{multline}
where $I$ is the identity matrix and the matrices $ N_{i} $ and $ P_{i} $
are respectively defined as follows:
\begin{equation}
\label{c_18}
	N_{i} = 
	\begin{pmatrix}
		0 & - {\tilde \omega}_{iz} & {\tilde \omega}_{iy} \\
		{\tilde \omega}_{iz} & 0 & - {\tilde \omega}_{ix} \\
		- {\tilde \omega}_{iy} & {\tilde \omega}_{ix} & 0 
	\end{pmatrix} \;\;\;\; i= 1,2 \; ,
\end{equation}
and 
\begin{equation}
\label{c_19}
	P_{i} = 
	\begin{pmatrix}
		{\tilde \omega}_{ix}^{2} & 
			{\tilde \omega}_{ix} {\tilde \omega}_{iy} & 
			{\tilde \omega}_{ix} {\tilde \omega}_{iz} \\
		{\tilde \omega}_{iy} {\tilde \omega}_{ix} & 
			{\tilde \omega}_{iy}^{2} & 
			{\tilde \omega}_{iy} {\tilde \omega}_{iz} \\
		{\tilde \omega}_{iz} {\tilde \omega}_{ix} & 
			{\tilde \omega}_{iz} {\tilde \omega}_{iy} & 
			{\tilde \omega}_{iz}^{2}
	\end{pmatrix} \;\;\;\; i= 1,2 \; .
\end{equation}
As we argued before, the very same classical propagator 
also maps the unit vectors 
$ {\bf n}_{1} $ and $ {\bf n}_{2} $, 
which identify an elliptic state, precisely into the new unit vectors 
$ {\bf n'}_{1} $ and $ {\bf n'}_{2} $  
of Eq. (\ref{c_12}), i.e. the unit vectors which label the 
elliptic state after it has evolved in time according to 
the quantum-mechanical propagator.

However, when ${\boldsymbol \omega}_{1}$ and ${\boldsymbol \omega}_{2}$
have constant orientation the dynamics can be understood more
intuitively by a {\it geometric} interpretation, as we illustrate 
more clearly in the next section. 

\section{excitation of circular states}

As a first application,
in this section we describe an alternative method for the excitation of
circular states, that is, hydrogenic states of maximum angular momentum.

Several diverse techniques have already been proposed and successfully 
implemented for the excitation of circular states and more
generally of large-{\bf L} elliptic states
{\cite{kleppner_83a,carlos_86b,delande_88a,spiess_93a,horsdal_94a}}.
However, all these methods are based on the {\it adiabatic} manipulation of the 
Rydberg electron wave function. First, the electron is 
excited to an eigenstate of the 
Hamiltonian of the hydrogen atom in weak fields, and next
the external fields are slowly varied in time while the electron always
remains {\it adiabatically} in the same eigenstate of the Hamiltonian. 
Therefore in all such techniques the time 
scale that defines the adiabatic regime is determined by 
the inverse of the spacing of the energy levels 
of the hydrogen atom in weak fields.
In practice, this means that a transformation is ``adiabatic"
if it takes place during a time much longer than the Stark or Larmor period
of the Rydberg electron.

However, ground state electrons are typically excited to high-{\it n} Rydberg states 
via a few optical transitions, and initially they are confined to
low angular momentum states. This causes some problems, because low-{\it l}
Rydberg electrons are strongly coupled to the atomic (or molecular) core, which 
enhances the probability of decaying out of the Rydberg state.
To the end of stabilizing the Rydberg electron it is then useful to
increase the angular momentum of the state as quickly as possible
{\cite{myself_14}}.
We propose a technique which is adiabatic with 
respect of the Kepler period of the electron, which is a much shorter
time than the Stark or Larmor periods (by a factor $\sim 1/n$). 
In fact, we do not try to maintain the electron in an eigenstate of the 
Hamiltonian at all times, and we only require that the dynamics must be 
confined within a hydrogenic {\it n}-manifold.

Our approach is based on the 
dynamics of elliptic states in weak external fields
(a method based on the same dynamics was 
suggested in Refs. {\cite{carlos_97c,nauenberg_94a}}),
and because we have shown that the evolution of these states is purely
classical, we can discuss the excitation of circular states
using classical mechanics. 
For a hydrogen atom in an electric field the red and blue Stark states 
with $m=0$ ($m$ being the usual magnetic quantum number) 
are two limit cases of elliptic states 
{\cite{delande_89a,delande_90a,delande_91a,nauenberg_89a,nauenberg_94a}}.
They correspond to two
classical ellipses with maximum eccentricity, which have collapsed to a 
straight line. 
Individual high-{\it n} Stark states can be accessed directly from 
low energy states via an optical transition, and we 
assume that the Rydberg electron is
initially placed in the blue Stark state, with $m=0$ 
(the same derivation 
applies also to electrons initially in the red Stark state).
The fields configuration under which the Stark state evolves into a circular state
could be derived analyzing the classical propagator of Eq. (\ref{c_17}), however
we present a more intuitive interpretation of the dynamics, which is based on a 
geometrical description of the time evolution.

\begin{figure}
\centerline{\psfig{file=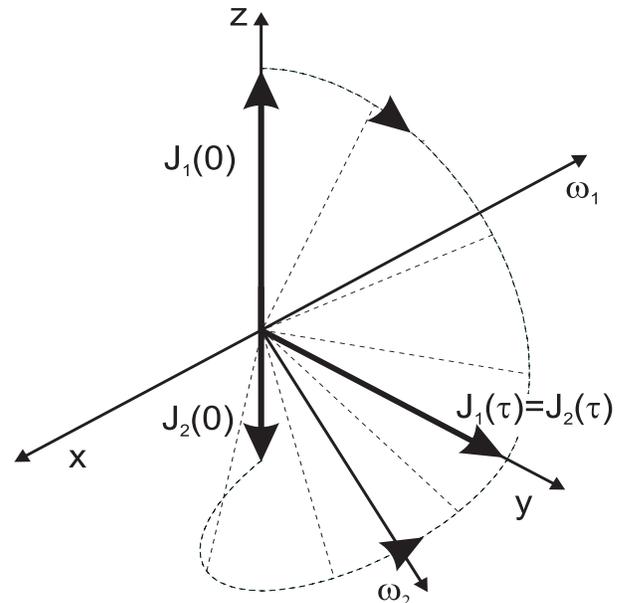,height=8.0cm,width=8.0cm,angle=0}}
\caption{
Dynamics in the rotating frame 
of the classical spin vectors ${\bf J}_{1}$ and ${\bf J}_{2}$
in an external field which satisfies the requirements described in 
the text, with $p=0$.
At time $t=0$ the two vectors point respectively along the 
$+z$ and $-z$ axis, which maximizes their difference, i.e. the 
Runge-Lenz vector, as it must be for the initial blue Stark state.
Next, they precess clockwise respectively around the 
two axes ${\boldsymbol \omega}_{1}$ and ${\boldsymbol \omega}_{2}$
(which lie in the {\it yz} plane),
and at the final time $\tau$ both vectors are aligned along the 
{\it y} axis. This means that the state has maximum angular momentum,
and it is the desired circular state.
The dashed curved lines show the trajectories described by the tips of
${\bf J}_{1}$ and ${\bf J}_{2}$, while the dashed straight lines represent the 
same vectors at some intermediate times. From the point of view shown in 
the figure, ${\bf J}_{1}$ precess {\it behind} its axis of rotation
${\boldsymbol \omega}_{1}$, whereas ${\bf J}_{2}$ passes {\it in front of}
${\boldsymbol \omega}_{2}$.
}
\label{c_fig_2}
\end{figure}

The external Stark field points along the positive {\it z} axis, and so does
the Runge-Lenz vector of the blue Stark state, which means that the two angular momenta
${\bf J}_{1}$ and ${\bf J}_{2}$ point respectively along the $+z$ and
$-z$ axis (see Fig. \ref{c_fig_2}). Clearly, the angular momentum ${\bf L}$
of the state vanishes, as it must
for an extreme Stark state. Our goal is to maximize ${\bf L}$, and therefore we 
need a configuration of external fields which will align ${\bf J}_{1}$ and 
${\bf J}_{2}$ so to maximize their sum (${\bf L} = {\bf J}_{1} + {\bf J}_{2}$)
and minimize their difference (${\bf a} = {\bf J}_{1} - {\bf J}_{2}$).
We construct such fields by rotating the Stark field counterclockwise (recall 
that a rotating electric field is equivalent to 
crossed electric and magnetic 
fields {\cite{goldstein_80,demkov_74,myself_7}})
around the {\it y} axis, and we also vary in time both the
magnitude of the field and its rotation frequency, that is:
\begin{multline}
\label{c_20}
	{\bf F} (t) = F(t) 
	\left[ 
		\cos \left( \int_{0}^{t} \omega_{R} (t') dt' \right) {\bf n}_{z} 
	\right. \\
	+ \left. 
		\sin \left( \int_{0}^{t} \omega_{R} (t') dt' \right) {\bf n}_{x} 
	\right] \; ,
\end{multline}
where ${\bf n}_{z}$ and ${\bf n}_{x}$ are respectively unit vectors along the 
{\it z} and {\it x} axis, and $\omega_{R} (t)$ is the time dependent rotation frequency. 
We also set some final time $\tau$, which must be long compared to the 
Kepler period to insure that the 
motion is confined within an {\it n}-manifold.
We require that at such time 
$\tau$ the field vanishes, so that the evolution of the elliptic state halts 
exactly when it becomes a circular state, that is,
\begin{equation}
\label{c_20a}
	F ( \tau ) = 0 \; .
\end{equation}
In other words, as we slowly rotate the Stark field we also slowly turn it off.

The effect of the field of Eq. (\ref{c_20}) is best analyzed in a frame 
rotating with the field.
A rotating frame is not a Galilean frame, 
and the inertial effects of the Coriolis forces can be 
described exactly by introducing an effective Larmor frequency equal to the 
rotation frequency ${\boldsymbol \omega}_{R}$ of the frame of reference
{\cite{goldstein_80,demkov_74,myself_7}}.
Therefore the equations of motion in the rotating frame are
\begin{equation}
\label{c_21}
	\begin{split}
		\frac{ d {\bf J}_{1} }{ dt } & =  
		- \left( {\boldsymbol \omega}_{R} + {\boldsymbol \omega}_{S} \right) 
		\times {\bf J}_{1} \\
		\frac{ d {\bf J}_{2} }{ dt } & =  
		- \left( {\boldsymbol \omega}_{R} - {\boldsymbol \omega}_{S} \right) 
		\times {\bf J}_{2} \; ,
	\end{split}
\end{equation}
where ${\boldsymbol \omega}_{R}$ and ${\boldsymbol \omega}_{S}$ 
point respectively along the {\it y} and {\it z} axis.
We then require that at all times
\begin{equation}
\label{c_22}
	\omega_{R} (t) = \omega_{S} (t) \; ,
\end{equation}
so that the axes of precession for ${\bf J}_{1}$ and ${\bf J}_{2}$ have 
constant orientation in space.
The axes of precession have
constant orientation in space also when 
the two frequencies are simply proportional to each other, but
the analysis of the dynamics is much simpler
when the Stark and rotation frequencies are exactly equal.

As we argued before, 
the two spin vectors precess around the two following ``magnetic fields":
\begin{equation}
\label{c_22a}
	\begin{split}
		{\boldsymbol \omega}_{1} & = 
		{\boldsymbol \omega}_{R} + 
		{\boldsymbol \omega}_{S} \\
		{\boldsymbol \omega}_{2} & = 
		{\boldsymbol \omega}_{R} - 
		{\boldsymbol \omega}_{S}  \; ,
	\end{split}
\end{equation}
where $ {\boldsymbol \omega}_{1} $ and $ {\boldsymbol \omega}_{2}  $
lie in the {\it yz} plane, and $ {\boldsymbol \omega}_{1} $ 
bisects the angle between the $+y$ and $+z$ axes,
whereas $ {\boldsymbol \omega}_{2}  $ bisects the angle between 
the $+y$ and $-z$ axes.

It is easy to see from Fig. \ref{c_fig_2} that a clockwise precession 
of ${\bf J}_{1}$ around ${\boldsymbol \omega}_{1}$ by an 
angle $\phi = \pi$ (or any odd integer multiple of $\pi$)
overlaps that spin vector with the $+y$ axis.
Similarly, a clockwise precession by the same angle and around 
${\boldsymbol \omega}_{2}$ aligns ${\bf J}_{2}$ along 
the $+y$ axis.
The net result of the time evolution is to align the two 
vectors with one another exactly, and so the blue Stark 
state evolves into the desired circular state, 
with angular momentum pointing 
along the $+y$ axis.
Therefore, we impose a final constraint on the 
total angle of precession, which translates to a condition on 
the magnitude of the Stark frequency of the external field:
\begin{equation}
\label{c_23}
	\int_{0}^{\tau} \sqrt{ \omega_{S}^{2} (t)  + \omega_{R}^{2} (t)  } dt =
	\sqrt{2} \int_{0}^{\tau} \omega_{S} (t)  dt = ( 2 p + 1 ) \pi \; ,
\end{equation}
where $p$ is some integer.
Equation (\ref{c_23}) also means that the total angle of rotation of 
the electric field is $ ( 2 p + 1 ) \pi / \sqrt{ 2 } $, which 
concludes our prescription for the excitation of circular states.

%
%
%
%
%
%
%

Finally, note that our analysis does not impose any constraint on ``how" the field is 
switched off. The only constraint is on the total angle of precession,
and the functional form of the time dependence of the field amplitude
may be chosen in the experimentally most convenient way {\cite{myself_14}}. 

\section{conclusions}

In this paper we have shown that the dynamics of quantum elliptic
states in weak external fields is described {\it exactly} by classical
perturbation theory. 
Therefore, the problem of evaluating a complicated quantum propagator is
reduced to the solution of the simple, linear
equations of motion of the classical system.
Clearly, in the case of fields with complicated time-dependence, 
one may have to solve the classical equations numerically, 
but that is still a relatively simple task.
Moreover, our work explains previous merely numerical
observations connection between classical and 
quantum dynamics; it also {\it generalizes to arbitrary fields} some 
theoretical arguments which were limited to some special 
configurations of the fields {\cite{carlos_97c,carlos_98a,nauenberg_94a}}.
Indeed, because of the properties of the coherent states of $ SO(4) $
{\cite{klauder_85,perelomov_86}}, 
our demonstration of the classical evolution of elliptic states in weak fields
holds for arbitrary fields
(although in this paper we did not solve such case analytically).
That is why our analysis provides a solid theoretical explanation 
for the surprising agreement between calculations
based on classical mechanics {\cite{myself_6,myself_7,myself_8,myself_9}} 
and several experimental results
{\cite{macadam_93a,gross_86a,gallagher_97}}.
It also indicates that it would be appropriate to use
perturbative, classical methods to analyze the dynamics of
Rydberg electrons in the complicated, time dependent fields that are 
expected under realistic ZEKE conditions {\cite{myself_14}}.

Atomic elliptic states ``sit" on classical
Kepler ellipses, and in a sense they {\it sew the wave flesh on the 
classical bones} {\cite{berry_72a}} made of periodic orbits.
Indeed, as the classical orbits slowly evolve in time
under the perturbation due to external weak fields,  
elliptic states follow exactly the same dynamics, and remain
confined along the very same ellipse throughout its motion.
Clearly, the argument can also be stated the other way around, and one 
may prefer to say that it is the classical orbit which is 
following the more fundamental quantum state.
Be it as it may, note that in the theory of atomic elliptic states there is 
no semiclassical approximation, and the correspondence with
classical mechanics is made directly from the purely quantum domain.

More technically, the dynamical equivalence between the motion of quantum elliptic
states and the time-averaged dynamics of classical orbits 
relies on the properties
of the coherent states of $SO(4)$, and on the 
fact that the external perturbations can be expressed in terms of the
generators of the group.
Our work then opens the question of the generality of our results.
That is: is the present example of quantum-classical equivalence a special
property of the Hydrogen atom only, or can it be extended to 
a wider class of weakly perturbed integrable systems?
This is a fundamental problem in modern physics, as 
it has been shown in the last few decades by the amount
of research on the quantum-to-classical
transition in nonintegrable systems {\cite{gutzwiller_90}}.

Finally, we have proposed an alternative, fast method for the 
production of ultrastable circular Rydberg states, which 
is based on the dynamics of atomic elliptic states.
In our derivation we make use of the exact quantum propagator 
for the purely hydrogenic Hamiltonian, which is only an approximation
to the case of more complex atoms. There, it is to be expected that 
the efficacy of the method may be partially spoiled by complex
core effects. 
It is likely that these effects are of minor magnitude 
and that they can be
compensated by a slight modification of the electric field,
or by the introduction of some magnetic field.
Our prescription provides then
a starting point for the search of the most effective fields
configuration, which can be reasonably expected 
to be ``in the neighborhood"
of the hydrogenic solution, and the tools of optimal control theory can
in principle be used to improve the efficiency of the method.
Further research in this area is currently in progress in our group.

\section*{Acknowledgments}

We wish to thank M. Nauenberg for useful comments
that helped us improve the clarity of our work.
This work was supported in part by NSF grant PHY94-15583 and
by the Army Research Office.

\bibliography{strings,myself,rydberg_1,rydberg_2,turgay,books,zeke}

\end{document}